\documentstyle[pra,multicol,aps]{revtex}
\def \epsi{\varepsilon}

\def \pad{\partial}
\def \bm#1{\mbox{\boldmath $#1$}}
\begin{document}

\setlength{\textwidth}{180mm}
\setlength{\textheight}{240mm}
\setlength{\parskip}{2mm}

\input{epsf.tex}
\epsfverbosetrue

\title{Optical Gap Solitons in Nonresonant Quadratic Media}

\author{Takeshi Iizuka$^{1,2}$ and Yuri S. Kivshar$^{2}$}

\address{$^1$ Department of Physics, Faculty of Science,
Ehime University, Ehime 790-8577, Japan \\
$^2$ Optical Sciences Centre, Australian National University, Canberra ACT
0200, Australia}

\maketitle
\begin{abstract}
We demonstrate an important role of the process of optical rectification in the
theory  of nonlinear wave propagation in quadratically nonlinear [or
$\chi^{(2)}$]  periodic optical media.  We derive {\em a novel physical model} for gap solitons in $\chi^{(2)}$ nonlinear Bragg gratings.
\end{abstract}

\pacs{PACS numbers: }

\vspace*{-1.0 cm}
\begin{multicols}{2}
\narrowtext

As has been recently demonstrated, large optical nonlinearities can be
generated in noncentrosymmetric media by means of the so-called {\em
cascading effects}, due to wave mixing under the condition of nearly
phase-matched second-harmonic generation  and other parametric processes
\cite{stegeman}. It has been also shown that
cascaded nonlinearities can support spatial optical solitons \cite{torner}
and  also different types of gap solitons in periodic Bragg gratings with a
quadratic [or $\chi^{(2)}$] nonlinear response \cite{gap}.

Importantly, when an input electromagnetic wave ${\bf E}$ at frequency
$\omega$ is launched into a noncentrosymmetric material, it generates also
a  quasi-static electric field (or {\em dc field}) at frequency zero. This
effect is known as {\em optical rectification}, and it is usually described
by a  contribution to the medium nonlinear polarization $P$ of the form
$P_i^0 =  \epsilon_0 \chi_{ijk}(0; \omega, -\omega) E_j(\omega)
E_k^*(\omega)$, where  $\chi_{ijk}(0; \omega, -\omega)$ is the nonlinear
optical susceptibility  describing optical rectification \cite{bass}. Such
an induced dc field changes a  refractive index via the linear
electro-optic effect. As has been recently shown
by Bosshard {\em et al.} \cite{gunter}, both theoretically and
experimentally,  the combined processes of optical rectification and the
linear electro-optic  effect lead to an {\em additional, nonresonant}
contribution into an effective  nonlinear refractive index of
noncentrosymmetric materials due to cascading  processes.

The effect of optical rectification is {\em usually neglected} in the
theory of quadratic solitons because the equation for the dc field can be integrated explicitly, leading to a nonresonant contribution into the effective cubic nonlinearity of the nonlinear Schr\"odinger (NLS) equation derived by means of the asymptotic technique in the  approximation of cascaded
nonlinearities (see, e.g., Ref. \cite{haus}). However,  for the propagation
of spatio-temporal multi-dimensional optical pulses in nonresonant
quadratic  media, such a reduction is no longer possible and, as a result,
the  multi-dimensional NLS equation becomes coupled to a dc field
\cite{ablowitz},  similarly to the integrable case
of the Dawey-Stewartson equation \cite{boiti}.

In this paper we show that the physical situation is qualitatively
different for periodic  quadratically nonlinear optical
media. We demonstrate that  coupling between the forward and backward waves
in one-dimensional shallow Bragg  gratings with a quadratic nonlinearity is
accompanied by a coupling to  the induced dc field that appears within the
same approximation and cannot be  eliminated by integration. {\em This
effect has been overlooked previously}, but it leads to {\em a novel
physical model for gap solitons} in quadratic media which we introduce
and analyze here.

We consider propagation of an optical pulse in a periodic medium with a
quadratic  $\chi^{(2)}$ nonlinear response. To derive the coupled-mode equations for the wave envelope, we start from Maxwell's equation,
\begin{equation}
c^2\nabla^2E-\frac{\pad^2}{\pad t^2}\left[ \hat\epsilon(z,i\pad_t)
+ \chi^{(2)}E\right] E =
0,
	\label{basic}
\end{equation}
where $\nabla^2$ stands for the Laplacian, $c$ is the speed
of light in vacuum, $E$ is the $x$-element of the  electric field, $\bm{E}=E(z,t)\bm{e_x}$, and  the quadratic nonlinearty is represented by a tensor element $\chi^{(2)}=\bm{\chi^{(2)}_{xxx}}$. We assume that $\hat\epsilon(z,\omega)$ is a periodic function of $z$, so it can be presented in a general form as a Fourier series,
\begin{equation}
\hat\epsilon(z,\omega)=
\epsilon(\omega)\left(1+\sum_{j=1}^\infty\epsilon_j e^{2ikz}
+\sum_{j=1}^\infty\epsilon_j^* e^{-2ikz}\right),
\label{grating}
\end{equation}
where $d =\pi/k$ is the period of the Bragg-grating structure.  Deriving the
couple-mode  equations below, we assume the case of a  {\em shallow
grating}, i.e. that the condition $\epsilon_j\ll1$ holds.  Additionally, we
may consider a periodic modulation of the nonlinear quadratic
susceptibility taking $\chi^{(2)}(z) = \chi^{(2)}(z+d)$ as a periodic
function  with the same period $d$.  However, we have verified that this
effect does not modify  qualitatively the analysis and results presented
below, so that we consider the simplest case when $\chi^{(2)}$ is constant.

For a periodic structure, the Bragg reflection leads to a
strong interaction between the forward and backward waves at the Bragg
wavenumber  $k_B \approx k$. To derive the coupled-mode equations for the
wave envelopes, we
consider the asymptotic expansion for the electric field in the form,
\begin{eqnarray}
&&E  = \left(E_+e^{ikz}+E_-e^{-ikz}\right)e^{-i\omega t}+ {\rm c.c.} \nonumber
\\
&& + E^{(0,0)}+E^{(0,2)}e^{2ikz}+E^{(0,-2)}e^{-2ikz} \nonumber \\
&& +\left(E^{(2,0)}\hskip -3pt+E^{(2,2)}e^{2ikz}\hskip
-3pt+E^{(2,-2)}e^{-2ikz}\right)
e^{-2i\omega t}  \hskip -3pt+{\rm c.c.},
\label{expa}
\end{eqnarray}
where $E_{\pm}= E_{\pm}(z,t)$ are slowly varying envelopes
of the forward(+) and bakward(-) waves. The frequency $\omega$ satisfies
the  dispersion relation for linear waves, $c^2k^2 =
\omega^2\epsilon(\omega)$. Due  to quadratic nonlinearity, the expansion
(\ref{expa}) includes  higher-order terms at the frequency $2\omega$ and
the zero-frequency term,  so that the slowly varying functions $E^{(n,m)}=
E^{(n,m)}(z,t)$ are defined as
nonlinear amplitudes of the $(n, m)$-order harmonics  $e^{-in\omega t}
e^{imkz}$.

Introducing a small parameter $\epsi$, we assume
$E_{\pm}\sim O(\epsi)$,  $\pad E_{\pm}/\pad t\sim \pad E_{\pm}/\pad z
\sim O(\epsi^3)$
and  $\epsilon_j \sim O(\epsi^2)$. Then, substituting the expansion
(\ref{grating}), (\ref{expa})
into Eq.  (\ref{basic}),  we compare the terms of
the same order in front of the coefficients $e^{-in\omega t}e^{imkz}$. At the
orders $(2,0)$, $(2,\pm 2)$ and $(0, \pm 2)$
 we respectively obtain,
\[
E^{(2,0)} = -\frac{2\chi^{(2)}}{\epsilon(2\omega)}E_+E_-\sim O(\epsi^2),
\]
\[
E^{(2, \pm2)} = -\frac{\omega^2 \chi^{(2)}}{[c^2k^2 - \omega^2
\epsilon(2\omega)]} E_\pm^2\sim O(\epsi^2),
\]
\[
E^{(0, \pm2)} = -\frac{\chi^{(2)}}{2c^2k^2}\frac{\pad^2}{\pad t^2}
\left(E_\pm^* E_\mp\right) \sim O(\epsi^6),
\]
where we have assumed non-resonant interaction with the second harmonic,
i.e.  $\omega^2\epsilon(2\omega)\neq c^2k^2$.
\par
At the orders $(1, \pm1)$ and $(0,0)$ we obtained a system of coupled
nonlinear
equations,
\begin{eqnarray}
&&i\left(\frac{\pad}{\pad t}+v_g\frac{\pad}{\pad z}\right)E_+
+\kappa E_-\nonumber\\
&&\qquad+\left(A|E_+|^2+B|E_-|^2+CE^{(0,0)}\right)E_+=0, \label{couple1}\\
&&i\left(\frac{\pad}{\pad t}-v_g\frac{\pad}{\pad z}\right)E_-
+\kappa^* E_+\nonumber\\
&&\qquad+\left(B|E_+|^2+A|E_-|^2+CE^{(0,0)}\right)E_-=0, \label{couple2}\\
&&\left(\frac{\pad^2}{\pad z^2} -\frac1{v_0^2}\frac{\pad^2}{\pad
t^2}\right)E^{(0,0)} \hskip -4pt + D \frac{\pad^2}{\pad t^2}\left(|E_+|^2
+|E_-|^2\right)=0,                                      \label{couple3}
\end{eqnarray}
where $v_g(\omega)=d\omega/dk$, $v_0=v_g(0)$,
$\kappa = \omega^2 \epsilon(\omega) \epsilon_1 f^{-1}(\omega)$,
$A=2(\chi^{(2)})^2\omega^4\{f(\omega)[c^2k^2-\omega^2\epsilon(2\omega)]\}^{-1}$,
$B=-4(\chi^{(2)})^2 \omega^2 [f(\omega) \epsilon(2\omega)]^{-1}$,
$C = 2\omega^2\chi^{(2)} f^{-1}(\omega)$, and $D =
-2\chi^{(2)}/c^2$ with
$f(\omega) \equiv [ \omega^2 \epsilon(\omega)]^{\prime}$.
If we keep the transverse coordinates ($x,y$), Eq. (\ref{couple3}) should
also include the transverse Laplacian in the same order.
System (\ref{couple1})-(\ref{couple3}) describes the interaction between
the forward and backward waves coupled to a dc wave induced via the
rectification  effect. Importantly, the order of the dc wave $E^{(0,0)}$ is
$\epsi^2$, so that  the dc field itself is of a higher order in
comparison with the forward and backward scattering waves. However, the dc field is coupled to  the fields $E_+$ and $E_-$ in the main order. Importantly, if we
assume much stronger dc field, e.g. of order of $O(1)$, the system is
decoupled and the  dc wave satisfies an independent equation.
Including the optical Kerr effect, we obtain the same equations
as Eqs. (\ref{couple1}), (\ref{couple2}) with the modified constants $A$ 
and $B$.

In the case of a single wave propagating in a homogeneous medium, the 
induced dc field is explicitly given by the host wave \cite{haus}. 
The similar result is valid
for the case of  a deep grating described by the  modulations of the Bloch
waves, but not for a shallow grating we discuss here. If we assume the
scaling $\pad E_{\pm}/\pad z \sim O(\epsi^2)$ (as usually done  in the
analysis of higher-dimensional systems such as the Dawey-Stewartson equation)
and $E^{(0,0)}$ of order $\epsi^4$,  then the coupling between the dc wave
and host-wave scan be neglected. However, for {\it isotropic scaling} as
presented  here, the effect of the dc wave is directly included into Eqs.
(\ref{couple1})  and (\ref{couple2}).  Interaction between the dc field and
fundamental harmonics has been also discussed in
Ref. \cite{haus}, however in that analysis, the dc field appears
as a cascading effect and its velocity is almost the same as the
phase velocity.  We notice that in our case, the dc field is essentially
excited  by quadratic nonlinearity and no assumption is required for the
velocity.

We are looking for spatially localized solutions of Eqs.
(\ref{couple1})-(\ref{couple3}) for {\em bright gap solitons} in the form,
\begin{eqnarray}
&&E_+ = \Delta^{-1/2}f(\zeta)e^{i[\theta_1(\zeta)-\Omega t+g/2]}, \nonumber \\
&&E_- = \Delta^{1/2} f(\zeta)e^{i[\theta_2(\zeta)-\Omega t-g/2]}.
\label{ansatz}
\end{eqnarray}
where $\zeta = z-Vt$;  the functions $f(\zeta)$ and
$\theta_{1,2}(\zeta)$,  and the parameters $\Omega$, $V$, $\Delta$
are assumed to be real. The parameter $g$  is the argument of the coupling
parameter $\kappa$, i.e. $\kappa = |\kappa| e^{ig}$. Substituting the
ansatz (\ref{ansatz}) into Eq. (\ref{couple3}),
we obtain
\[
E^{(0,0)}(\zeta) = -\frac{v_0^2V^2 D}{(v^2_0-V^2)}\left(\Delta +
\frac{1}{\Delta} \right) f^2(\zeta),
\]
and therefore the contribution of the dc field should vanish at $V=0$.

From Eqs. (\ref{couple1}) and (\ref{couple2}), we set the parameter
$\Delta$  as $\sqrt{(v_g - V)/(v_g + V)}$,  and then obtain a system of
coupled  equations for $f$, $\theta _- \equiv \theta_1 - \theta_2$ and
$\theta_+ \equiv \theta_1 +\theta_2$,
\begin{eqnarray}
&&\frac{df}{d\zeta }=\mu f \sin \theta_-,             \label{fphi1}\\
&&\frac{d \theta_-}{d\zeta }  + \nu - 2\mu\cos \, \theta_-
+\delta f^2=0,
\label{fphi2}\\
&&\frac{d \theta_+ }{d\zeta}  + \frac{V \nu}{v_g} + \eta f^2=0,
\label{thetawa}
\end{eqnarray}
where $\quad\mu = |\kappa|/ (v_g^2-V^2)^{1/2}$,
$\nu=-(2v_g \Omega)/(v_g^2-V^2)$,
and
\begin{eqnarray*}
&&\delta=-2(v_g^2-V^2)^{-1/2}\left(\frac{(v_g^2+V^2)}{(v_g^2-V^2)}\tilde{A}
+ \tilde{B} \right),
\\
&&\eta=-4(v_g^2-V^2)^{-3/2}v_gV \tilde{A}, \\
&& \tilde{A} =A-\frac{V^2v_0^2CD}{(v_0^2-V^2)}, \qquad
\tilde{B} = B-\frac{V^2v_0^2CD}{(v_0^2-V^2)}.
\end{eqnarray*}
We notice that in the problem under consideration, we should assume
$|V|<v_g$ and $|\nu|< 2\mu$.

Similar to the analysis presented in Ref. \cite{mills}, from Eqs. (\ref{fphi1})-(\ref{thetawa}) we obtain a closed  differential equation for the function  $\theta_-$ in the form of the double sine-Gordon (DSG) equation. The DSG equation can be integrated, and its localized solutions  are two types of {\em kinks} and {\em anti-kinks} \cite{braun}. Using the relevant solutions, we can then find
\[
f(\zeta) = \left\{ \frac{\mp (4\mu / \delta)[1- ( \nu / 2\mu )^2 ]}
{\cosh (\zeta \sqrt{4\mu^2-\nu^2}) \mp (\nu / 2\mu) } \right\}^{1/2},
\]
where the signs $\pm$ stand for the cases $\delta>0$ and $\delta<0$,
respectively.
Functions $\theta_1$ and $\theta_2$ are then obtained as
\begin{eqnarray}
&&\theta_- = \theta_1-\theta_2 \nonumber\\
&&=-2 \tan^{-1} \left\{\sqrt{\frac{2\mu-\nu}{2\mu+\nu}}
 \; \tanh ^{\mp1} \left(\frac{\sqrt{4\mu^2-\nu^2}}2\zeta
\right)^{\mp1}\right\},
\label{hiku}\\
&&\theta_+ = \theta_1+\theta_2 = -\frac{\nu V}{v_q}\zeta\nonumber\\
&&\pm\frac{4\eta v_g}\delta
\tan^{-1} \left[ \sqrt{\frac{2\mu\pm\nu}{2\mu\mp\nu}}
\tanh \left( \frac{\sqrt{4\mu^2-\nu^2}}2\zeta \right)\right]
+{\cal C}_{\pm},
\label{add}
\end{eqnarray}
where ${\cal C}_{\pm}$ are integration constants. In order to obtain solutions for gap solitons, we restrict the possible angle variable by the  domain, $0
\le \theta_- \le 2\pi$. The solution obtained from Eq.
(\ref{hiku}) describes a two-parameter family of gap solitons, spatially
localized waves in the Bragg gratings, which are similar to the gap solitons
of the conventional  couple-mode theory.  Actually, by renormalizing the
variables as  $A\to\pm\sigma$, $B\to\pm1$, $C\to0$, $|\kappa|\to1$, $\qquad
v_g\to1$, $V\to
v$, $\Omega\to\pm\sqrt{1-v^2}\cos Q$, ${\cal C}_{\pm} \to (1/2\pm1/2)\pi \pm
(4\sigma v\alpha^2)/(1-v^2)\pi+2\phi$,  we can demonstrate that the
solution is essentially the same as that earlier obtained  in Ref. \cite{aceves}. However, the effect of the dc wave is included in the parameters  $\delta$ and $\eta$.

In the case $|\nu|>2\mu$, spatially localized solution of Eqs. (\ref{fphi1}), (\ref{fphi2}) do not exist.  Instead,  the kinks of the DSG equation for $\theta_-$  give solutions for {\em dark gap solitons} (see also \cite{feng}), localized waves on nonvanishing backgrounds, 
\[
f(\zeta)=\sqrt{\frac{2\mu}{|\delta|}\left(\frac{|\nu|}{2\mu}-1\right)}
\frac{ \sqrt{|\nu|/2\mu} \cosh (\sigma\zeta) \pm1 }
{ \sqrt{(|\nu|/2\mu) \cosh^2 (\sigma\zeta)-1}}, 
\]
where
\[
\sigma = 4\mu \sqrt{\frac{|\nu|}{2\mu}-1}.
\]
Upper and lower signs correspond to two types of such solitons, with the maximum intensity {\em large} or {\em smaller} than the background intensity.

The effective renormalisation of the coefficients due to the induced dc field seems extremely important for the soliton stability. Indeed, when $v_0 < v_g$ the coefficients have a singularity provided $V \rightarrow  v_0$,  changing the character of the dependence of the soliton parameters and the system conserved quantities on $V$. The recent stability analysis of the conventional gap solitons \cite{stability} revealed the existence of two types of instabilities, {\em oscillatory} and {\em translational}. The most important, translational
instability appears for large $V$, so that the induced dc field is expected
to have a strong effect on the soliton stability.  In fact, we anticipate that all gap solitons for $ v_0 < V < v_g$ may become unstable, similar to the solitary  waves in hydrogen-bonded molecular systems \cite{zolo}. It is also worth to mention that in the limit $V \to v_0$ when the coefficients grow,  
the transverse effect in Eq. (\ref{couple3}) become important  and should be included to compensate the singularity.

To demonstrate that the issue of the soliton stability becomes nontrivial
for this model, we present the system invariants. Similar to some other
models, we are not able to present Eqs. (\ref{couple1})-(\ref{couple3}) in
a Hamiltonian form directly, and therefore we introduce an auxiliary
function $\phi$ through the relation, $\alpha \pad \phi/\pad
z=E^{(0,0)}-v_0^2D(|E_+|+|E_-|^2)$, where $\alpha^2=v_0^2D/C$. Then,
we define the second canonical variable as $\psi = v_0^2 \partial \phi/\partial t$ and show that Eqs. (\ref{couple1})-(\ref{couple3}) can be written 
as a Hamiltonian system
\[
\frac{\pad \phi}{\pad t} = \frac{\delta H}{\delta \psi}, \;\;\;
\frac{\pad \psi}{\pad t} = -\frac{\delta H}{\delta \phi}, \;\;\;
\frac{\pad E_{\pm}}{\pad t} = i\frac{\delta H}{\delta E^*_{\pm}},
\]
with the following Hamiltonian,
\begin{eqnarray}
&&H=\int_{-\infty}^{+\infty} dz\left\{
\frac{v_0^2}2\psi^2-\phi\frac{\pad^2\phi}{\pad z^2}
+\kappa E_+^*E_-+\kappa^* E_+E_-^*
\right.\nonumber\\
&&\ \left.+\frac{iv_g}{2} \left(E_+^*\frac{\pad E_+}{\pad z}
-E_+\frac{\pad E_+^*}{\pad z}
-E_-^*\frac{\pad E_-}{\pad z}+E_-\frac{\pad E_-^*}{\pad z}\right)
\right.\nonumber\\
&&\ \left.+\frac{\bar A}2(|E_+|^4+|E_-|^4)
+\bar B|E_-E_+|^2\right.\nonumber\\
&&\ \left.+\alpha C\frac{\pad \phi}{\pad
z}\left(|E_+|^2+|E_-|^2\right)\right\},
 \label{H}
\end{eqnarray}
where $\bar A=A+v_0^2CD$ and $\bar B=B+v_0^2CD$.
Other integrals of motion of the system (\ref{couple1})-(\ref{couple3}) are
the field momentum,
\[
P = \int_{-\infty}^{+\infty} dz \left\{ \left(E_+\frac{\pad	E_+^*}{\pad z}
+ E_-\frac{\pad E_-^*}{\pad z}\right) - 2\frac{\pad\phi}{\pad z}\psi \right\},
\]
 the total number of the forward and backward waves, and an independently conserved number of the dc waves,
\[
N=\int_{-\infty}^{+\infty} dz(|E_+|^2+|E_-|^2), \;\;
N_0=\int_{-\infty}^{+\infty} dz\psi.
\]
Therefore, in a sharp contrast to the conventional theory of gap solitons,
the model (\ref{couple1})-(\ref{couple3}) possesses one additional integral
of motion, it has no analogy with other soliton-bearing nonintegrable models
where the soliton stability has been investigated so far.

In conclusion, we have demonstrated that in periodic optical media with a
quadratic nonlinear response, Bragg-grating-induced  coupling between the
forward and backward propagating modes leads always to an induced dc field which plays an important role for nonlinear pulse propagation in periodic  $\chi^{(2)}$ media.  We have derived a novel model of the coupled-mode theory for optical gap solitons in quadratically nonlinear Bragg gratings, and have found the analytical solutions for moving bright and dark gap  solitons.  Analysis of the soliton stability for the model (\ref{couple1})-(\ref{couple3})
will be presented elsewhere.

Takeshi Iizuka acknowledges a hospitality of the Optical Sciences Centre
and a support of the Japanese Ministry of Education. Yuri  Kivshar is a member of the Australian Photonics Cooperative Research Centre.

\end{multicols}
\end{document}